\documentclass[twocolumn,showpacs,prb]{revtex4}
\usepackage{amsfonts}
\usepackage{amsmath}
\usepackage{amssymb}
\usepackage{graphicx}

\begin{document}

\title{Tuning of energy levels and optical properties of graphene quantum
dots}
\author{Z. Z. Zhang}
\author{Kai Chang}
\altaffiliation[Corresponding author:]{kchang@red.semi.ac.cn}
\affiliation{SKLSM, Institute of Semiconductors, Chinese Academy of Sciences, P. O. Box
912, Beijing 100083, China}
\author{F. M. Peeters}
\affiliation{Departement Fysica, Universiteit Antwerpen, Groenenborgerlaan 171, B-2020
Antwerpen, Belgium }

\begin{abstract}
We investigate theoretically the magnetic levels and optical properties of
zigzag- and armchair-edged hexagonal graphene quantum dots (GQDs) utilizing
the tight-binding method. A new bound edge state at zero energy appears for
the zigzag GQDs in the absence of a magnetic field. The magnetic levels of
GQDs exhibit a Hofstadter-butterfly spectrum and approach the Landau levels
of two-dimensional graphene as the magnetic field increases. The optical
properties are tuned by the size, the type of the edge, and the external
magnetic field.
\end{abstract}

\maketitle

Graphene is a single atomic layer consisting of a two-dimensional honeycomb
lattice of carbon atoms. This novel system has attracted intense attention
because of new fundamental physics and promising applications in
nanoelectronics \cite{Novoselov,Novoselov2}. It exhibits high crystal
quality, an exotic Dirac-type spectrum, and ballistic transport properties
on a submicron scale. Graphene samples are usually fabricated by
micromechnical cleavage of graphite and have excellent mechanical properties
that make it possible to sustain huge electric currents. The lateral
confinement of Dirac fermions in graphene is still an enigmatic and
extremely challenging task due to the well-known Klein paradox. The Klein
paradox makes it impossible to localize the carriers in a confined region
utilizing an electrostatic gate. The confinement of Dirac fermions at a
nanometer scale is one of the central goals of graphene-based electronics
and has attracted increasing interest\cite%
{Efetov,Egger,HChen,Peeters,Trauzettel,Antidot}. Recently it was
demonstrated
experimentally that graphene can be cut in the desired shape and size\cite%
{Novoselov,Novoselov2}. Recent progresses in fabricating and
characterizing stable graphene nanostructures provides the
opportunity to explore the various remarkable
optical\cite{Falko,Carbotte,Reichl} and transport
properties\cite{Kim} of these structures.

\begin{figure}[tbp]
\includegraphics [width=0.9\columnwidth]{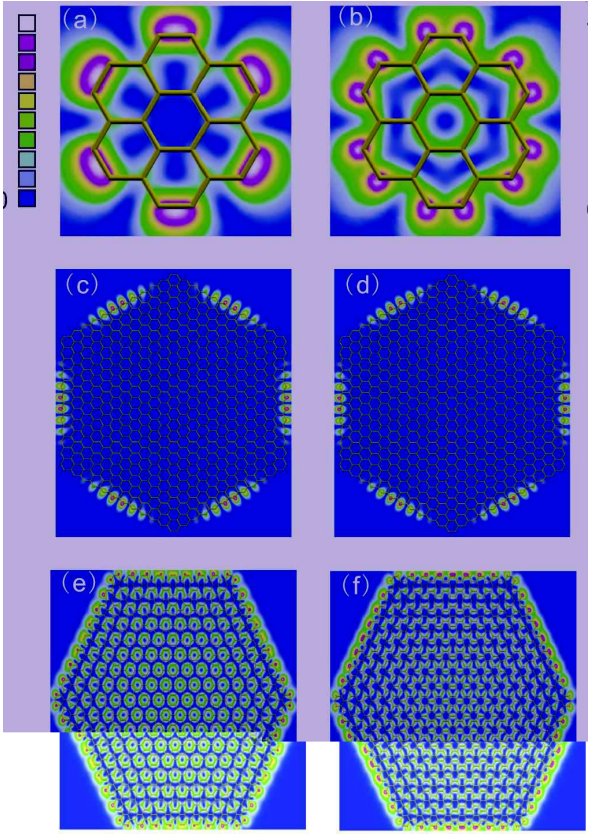}
\caption{ (Color online) Electronic density distributions of the highest
valence level (HVL) and lowest conduction level (LCL) in the absence of
magnetic field. Panels (a) and (b): HVL and LCL for the $N_z=2$ zigzag-edged
graphene quantum dot (ZGQD). Panels (c) and (d) : the same with $N_{z}=12$.
Panels (e) and (f): the same with armchair-edged graphene quantum dot
(AGQD). }
\label{fig1}
\end{figure}

In this work, we investigate theoretically the electronic structure and
optical properties of zigzag- and armchair-edged hexagonal graphene quantum
dots (GQDs) (see Fig. \ref{fig1}) utilizing the nearest-neighbor
tight-binding model. The dangling bonds at the edges are passivated by
hydrogen atoms. The model has been successfully used for fullerene
molecules, carbon naotubes, and other carbon-related materials \cite%
{Nakada,Waka,Saito,Ezawa}. The Hamiltonian of GQDs can be written as $%
H=\sum\limits_{i}\varepsilon _{i}c_{i}^{\dag
}c_{i}+\sum\limits_{\left\langle i,j\right\rangle }t_{i,j}c_{i}^{\dag }c_{j}$%
, where $\varepsilon _{i}$ is the site energy, $t_{ij}$ is the transfer
energy between the nearest-neighbor sites, and $c_{i}^{\dag }$ ($c_{i}$) is
the creation (annihilation) operator of the $\pi $ electron at the site $i$.
When considering a magnetic field $B$ applied perpendicularly to the plane
of a GQD, the transfer integral $t_{ij}$ becomes $t_{ij}=te^{i2\pi \phi
_{i,j}},$ where $\phi _{ij}=\frac{e}{h}\int_{r_{i}}^{r_{j}}d\mathbf{l}\cdot
\mathbf{A}$ is the Peierls phase. $\mathbf{A}=\left( 0,Bx,0\right) $ is the
vector potential corresponding to the magnetic field B along the \textit{z}
axis, which is perpendicular to the graphene plane. In our calculation, we
take $\Phi _{0}=h/e$ as the unit of the magnetic flux and $\Phi =\sqrt{3}%
Ba_{0}^{2}/2$ as the magnetic flux through a plaquette, where $a_{0}=2.46%
\mathring{A}$ is the lattice constant of graphite. The difference between
the values of $\varepsilon _{i}$ and $t_{ij}$ for the atoms at the edge and
the center is neglected. The relevant parameters used in our calculation are
$\varepsilon =0$, $t=-3.033$\cite{Saito}. The eigenvalues and eigenstates
can be obtained from the secular equation $\det \left\vert \varepsilon
-H\right\vert =0$, where $H_{ii}=0,H_{\left\langle i,j\right\rangle
}=te^{i2\pi \phi _{i,j}}$.

Figs. \ref{fig1} show electronic density distributions of the zigzag and
armchair-edged graphene quantum dots (ZGQD and AGQD, respectively), in the
absence of a magnetic field. The size of a dot is characterized by N, the
number of hexagonal units along an edge. Figure 1(a) and 1(b) show the
probability distributions of the highest valence level (HVL) and lowest
conduction level (LCL) for the ZGQD with small size ($N=2$). The probability
distributions of the HVL and LCL correspond to the bonding and anti-bonding
states that are localized at the corner of the hexagonal GQD. In contrast to
conventional semiconductor quantum dots where the ground state is localized
at the center of dot, the ground states for the conduction and valence
bands, i.e., HVL and LCL, localize at the middle of each edge in the
ZGQD(see Fig. \ref{fig1}(c) and (d)) as the size of the ZGQD increases. This
feature can be understood as follows: the Dirac fermion in a ZGQD behaves
like a confined photon in a cavity, and the lowest mode is the whispering
gallery mode, which also localizes at the boundary of the cavity. The
difference between the bonding and anti-bonding states becomes smaller as
the size of the ZGQD increases. The difference between the edge states of a
ZGQD and a graphene nanoribbon is that the edge state of the ZGQD localizes
at the middle of the edge of GQD, in contrast to the homogeneously
distributed edge state of a zigzag graphene nanoribbon\cite{Waka,Brey} or a
zigzag triangular GQD\cite{Yamamoto}. This occurs because the contribution
of each carbon atom at the edge of a ZGQD to the edge state is different,
while it is the same for a zigzag nanoribbon or a zigzag triangular GQD. The
density distributions of the LCL\ and HVL in an AGQD (see Figs. \ref{fig1}%
(e) and (f)) extend more completely over the whole GQD region and are very
different from that in a ZGQD.\ This difference is indeed caused by the
different topological geometry of the boundary of the graphene
nanostructures.

\begin{figure}[tbp]
\includegraphics [width=\columnwidth]{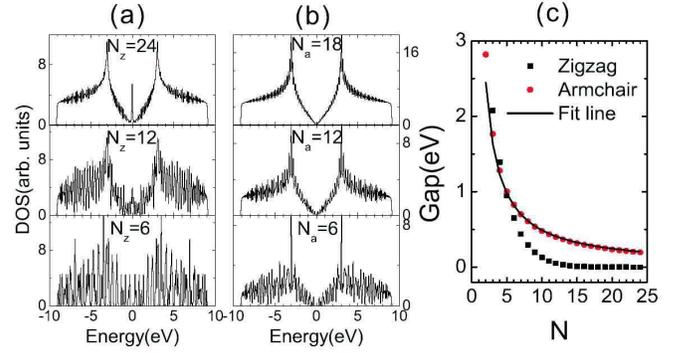}
\caption{(Color online) Density of states of ZGQD (a) and AGQD (b). We use a
Gaussian function $f(E)=e^{-(E-E_{0})^{2}/\Gamma ^{2}}$ with a broadening
factor $\Gamma =0.05$ eV to smooth the discontinuous energy spectra. (c) The
gap of ZGQD and AGQD as a function of the size, and the function of the fit
line is $a/N$ with $a=4.9$ eV.}
\label{fig2}
\end{figure}

Figs. \ref{fig2}(a) and (b) show the density of states (DOS) of ZGQDs and
AGQDs, respectively, with different sizes in the absence of a magnetic
field. The total number of the carbon atoms in ZGQD and AGQD are $6N_{z}^{2}$
and $6(3N_{a}^{2}-3N_{a}+1)$, respectively. From the figures, we find that
there is no edge state in a small ZGQD, and the edge state appears when the
size of ZGQD\ increases according to the states at zero energy. Meanwhile,
there is never an edge state for the AGQD. To demonstrate how the edge state
appears, we plot the energy gap, i.e., the energy difference between the
lowest conduction band level (LCL) and the highest valence band level (HVL),
as a function of the size ($N$) of the GQD in Fig. \ref{fig2}(c). The energy
gap decreases as the size of the GQD increases. Interestingly, the energy
gap of the zigzag (armchair) GQD decays to zero quickly(slowly) as the size
of the GQD increases. When the size of the AGQD approaches infinity, the gap
decreases to zero, i.e., we recover the two-dimensional graphene case. The
calculated energy gap for the AGQD falls off as $1/N$ $\propto 1/L$ (see the
solid line in Fig. \ref{fig2}(c)), where $L$ is the length of each edge of
the hexagonal GQD. This dependence of the band gap on the size of GQD is
very different from that of a conventional semiconductor QD, which behaves
as $1/L^{2}$.

\begin{figure}[tbp]
\includegraphics [width=\columnwidth]{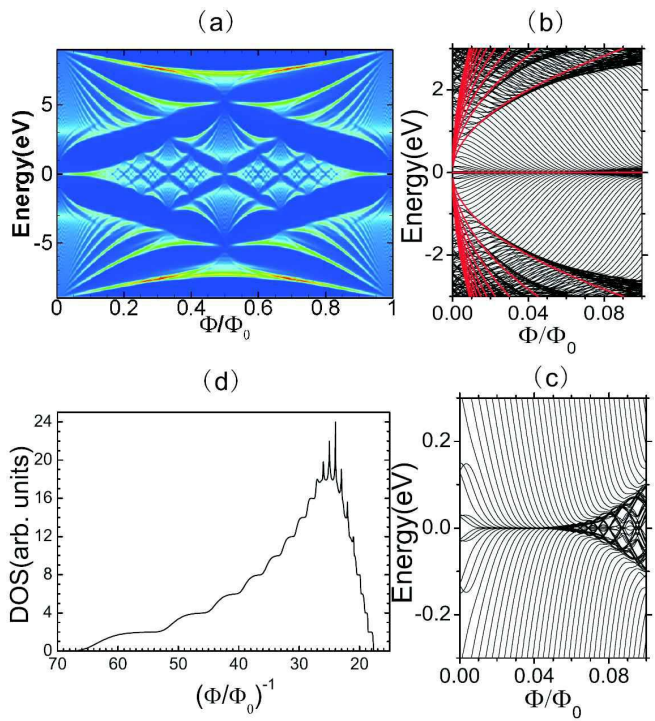}
\caption{ (Color online) (a) The spectrum of the $N_{z}=12$ ZGQD in a
magnetic field. We use a Gauss function with a broadening factor of 0.1 eV
to smooth discontinuous energy spectra. (b) and (c) the magnetic energy
level fan near the Dirac point, i.e., the zero energy point. The red lines
in (b) correspond to the Landau level of two-dimensional graphene. (d) the
DOS at the Dirac point as a function of the inverse flux $\Phi /\Phi _{0}$,
where we use a Gauss function with a small broadening factor of 0.01 meV.}
\label{fig3}
\end{figure}

\begin{figure}[tbp]
\includegraphics [width=\columnwidth]{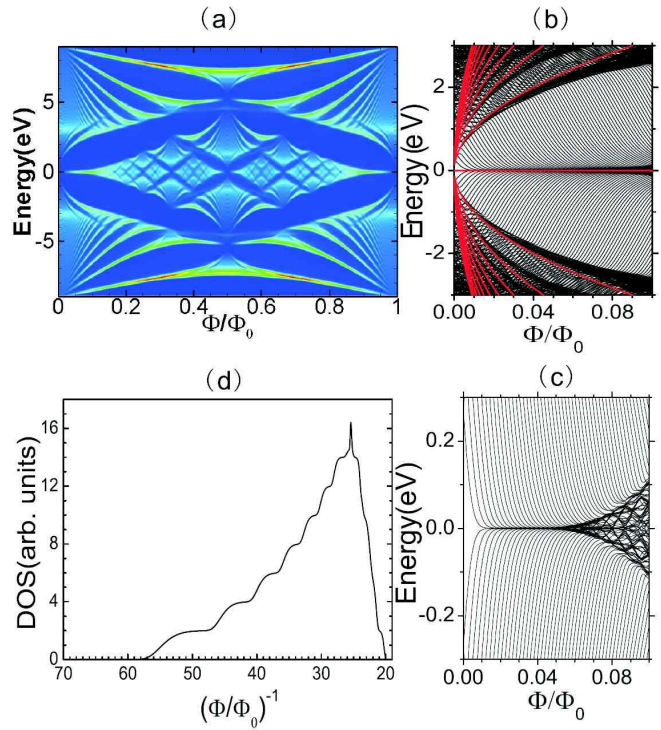}
\caption{ (Color online) The same as Fig.3, but for the $N_{a}=9$ AGQD}
\label{fig4}
\end{figure}

Figs. \ref{fig3}(a) and \ref{fig4}(a) depict the magnetic field dependence
of the energy spectrum of a ZGQD and AGQD exhibiting a clear Hofstadter
butterfly characteristic, which is fractal and exhibits self-similarity\cite%
{Hosfstadter,Waka,Aoki,Nemec}. As the magnetic flux increases, the
magnetic levels in the GQD, i.e., the so-called Fock-Darwin levels,
approach the Landau levels (see
the red lines in Figs. \ref{fig3}(b) and \ref{fig4}(b)) in graphene $%
E_{n}=sgn\left( n\right) (\sqrt{3}ta_{0}/2l_{B})\sqrt{2\left\vert
n\right\vert }$, where $l_{B}=\sqrt{\hbar /eB}$ is the cyclotron radius, $n$
is an integer, and $sgn$ is the sign function.

Figs. \ref{fig3}(c) and \ref{fig4}(c) show in detail how the magnetic levels
of ZGQD and AGQD approach the zero-th Landau level at small magnetic flux.
As magnetic flux increases, more energy levels approach the zero-th Landau
level in pairs. The degeneracy of the energy level at zero energy will reach
its maximum value $2N$ for $\Phi /\Phi _{0}=1/2N$. When the magnetic flux
increases further, the degeneracy of the energy level at zero energy is
lifted fast. This feature can also be seen in Fig. \ref{fig3}(d), which
plots the DOS\ at the Dirac point. This figure indicates that the
degeneracy, i.e., the number of energy levels at the zero energy, \emph{%
approximately} decreases inverse linearly with the magnetic flux $\Phi /\Phi
_{0}$. These figures clearly demonstrate that the energy spectrum of the GQD
possesses electron-hole symmetry when we neglect the second-nearest-neighbor
interaction. The DOS and the magnetic level fan of the AGQD are similar to
that of the ZGQD except at small magnetic flux. Comparing Fig. \ref{fig3}(c)
to Fig. \ref{fig4}(c), the magnetic levels in the AGQD are distinct from
those in the ZGQD at small magnetic flux, because the ZGQD shows the edge
state and AGQD does not for the levels near the Dirac point in absense of
magnetic field. Therefore, the magnetic levels exhibit distinct behavior as
the magnetic flux increases. The DOS of the AGQD (see Fig. \ref{fig4}(d))
also shows a step-like feature as the magnetic flux at the Dirac point
increases.

\begin{figure}[tbp]
\includegraphics [width=\columnwidth]{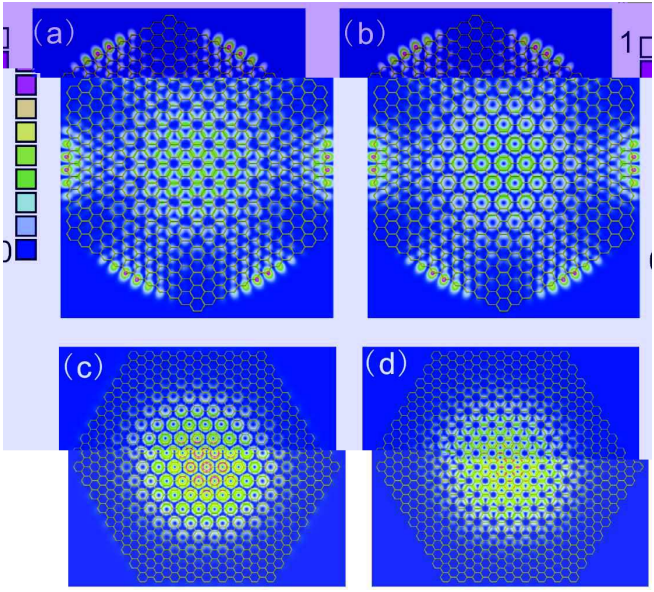}
\caption{ (Color online) (a) and (b) show the density distributions of the
HVL and LCL for the $N_{z}=12$ ZGQD in the presence of the magnetic flex $%
\Phi /\Phi _{0}=0.01$, respectively. (c) and (d) the same as (a) and (b),
but now for $N_{a}=9$ AGQD.}
\label{fig5}
\end{figure}

\begin{figure}[tbp]
\includegraphics [width=\columnwidth]{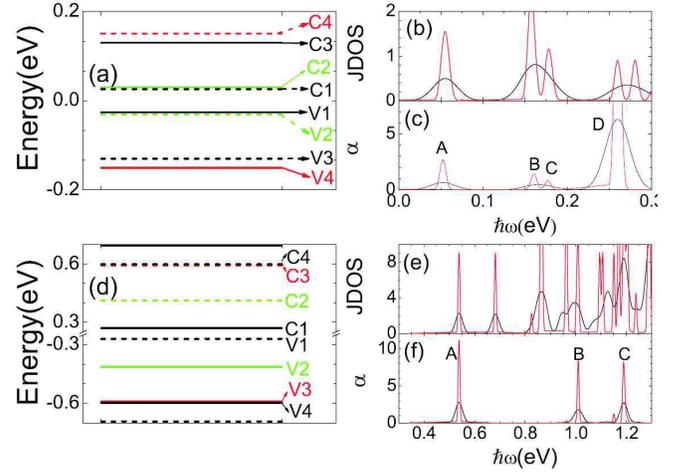}
\caption{ (Color online) (a) and (d) are the level diagram for $N_{z}=12$
ZGQD and $N_{a}=9$ AGQD without the magnetic field, where different
symmetries are represented by different colors and lines: black solid, black
dashed, red solid, red dashed, green solid, and green dashed lines for the $%
E_{1}$, $E_{2}$, $A_{1}$, $A_{2}$, $A_{3}$, and $A_{4}$ irreducible
representations of the $C_{6v}$ symmetry, respectively. (b) and (c) the JDOS
and the optical absorption spectrum $\protect\alpha $ for $N_{z}=12$ ZGQD.
We used a Gauss function with different broadening factors: 0.02 and 0.005
eV for the black and red line. (e) and (f) are the same as (b) and (c), but
for $N_{a}=9$ AGQD.}
\label{fig6}
\end{figure}

Fig. \ref{fig5} describes the density distributions of the LCL and HVL in
the ZGQD and AGQD at small magnetic flux $\Phi /\Phi _{0}=0.01$.
Interestingly, the density distributions of the LCL and HVL penetrate into
the center of the GQD for the ZGQDs, which is very different from the AGQD
case where both the electron and hole are dominantly localized in the center
of the GQD. The density distributions for the ZGQD and AGQD show $C_{6v}$
symmetry. This characteristic is caused by the magnetic confinement when the
magnetic length $l_{B}$ becomes comparable with the size of the GQD. In
addition to those differences, the LCL and HVL of the zigzag GQD show
opposite symmetry order with respect to that of the armchair GQD, i.e., the
LCL (HVL) and HVL(LCL) of the ZGQD (AGQD) belong to the $E_{1}$($E_{2}$) and
$E_{2}$($E_{1}$) representations at zero magnetic field (see Fig. \ref{fig6}%
).

The optical properties of GQDs are promising for potential applications in
optic-electronic devices based on graphene. Therefore, we calculate the
absorption spectra of GQD $\alpha (\hslash \omega )=\frac{\pi e^{2}}{%
m_{0}^{2}\varepsilon _{0}cn\omega V}\sum\limits_{c,v}|\vec{\varepsilon}\cdot
P_{cv}|^{2}\times \delta \left( E_{c}-E_{v}-\hbar \omega \right) ,$ where $n$
is the refractive index, $c$ the speed of light in vacuum, $\varepsilon _{0}$
the permittivity of vacuum, $m_{0}$ the free-electron mass, and $\vec{%
\varepsilon}$ is the polarization vector of the incident light along the $x$
direction. The coupling between the $sp_{2}$ states and the $p_{z}$ state is
neglected since we are only interested in the optical properties of the GQD
near the Dirac point, i.e., at the low energy regime. The momentum matrix%
\cite{Pedersen} is $\left\langle n\right\vert \mathbf{p}\left\vert
m\right\rangle =im_{0}/\hbar \sum\limits_{\mathbf{r}}\sum\limits_{\mathbf{r}%
^{\prime }}c_{\mathbf{r}}^{\ast }c_{\mathbf{r}^{\prime }}\left( \mathbf{r}%
^{\prime }-\mathbf{r}\right) \left\langle p_{z},\mathbf{r}\right\vert
H\left\vert p_{z},\mathbf{r}^{\prime }\right\rangle $. The momentum operator
$p_{x}(p_{y})$ has $E_{2}$ symmetry and its direct product with all the
irreducible representations of the C$_{6v}$ group can be found in Table \ref%
{table1}. We divide the levels of the GQD into two different families: $%
A_{1},A_{3},E_{1}\in \Omega _{1}$ and $A_{2},A_{4},E_{2}\in \Omega _{2}$.
The symmetry requires that only transitions between the valence band levels
and the conduction band levels belonging to the different families $\Omega
_{1}$ and $\Omega _{2}$ are allowed. Notice that the initial or final states
of the transition should belong to the $E_{1}$ or $E_{2}$ representations.
\begin{table}[tbp]
\caption{ Direct products of the $E_{2}$ representation for momentum
operator $p_{x}(p_{y})$ with all the irreducible representations of the $%
C_{6}v$ group. The results are presented as direct sums of all possible
irreducible representations of the $C_{6}v$ group. The notations of
symmetries are adopted from Ref. \onlinecite{Group}}
\label{table1}%
\begin{ruledtabular}
\begin{tabular*}{\columnwidth}{cc}
Direct product & Direct sum \\
\hline
$E_2$$\otimes$$A_1$ & $E_2$\\
$E_2$$\otimes$$A_2$ & $E_1$\\
$E_2$$\otimes$$A_3$ & $E_2$\\
$E_2$$\otimes$$A_4$ & $E_1$\\
$E_2$$\otimes$$E_1$ & $A_2$$\oplus$$A_4$$\oplus$$E_2$\\
$E_2$$\otimes$$E_2$ & $A_1$$\oplus$$A_3$$\oplus$$E_1$\\
\end{tabular*}%
\end{ruledtabular}
\end{table}
In Fig. \ref{fig6}(a) and (d), we label the level structure of a $N_{z}=12$
and $N_{a}=9$ GQD near the Dirac point as $C_{1}-C_{n}$ for conduction bands
with ascending order and $V_{1}-V_{n}$ for valence bands with descending
order, respectively. The conduction band levels $C_{i}$ and valence band
levels $V_{i}$\ belong to the distinct families $\Omega _{1}$ and $\Omega
_{2}$, respectively. For example, if $C_{i}$ belongs to the family $\Omega
_{1}$, i.e., $A_{1}$, $A_{3}$ or $E_{1}$, $V_{i}$ must belong to the family $%
\Omega _{2}$, i.e., $A_{2},$ $A_{4}$ or $E_{2}$, \textit{or} \textit{vice
versa}. For zigzag GQDs with even $N_{z}$, the conduction band levels, from
bottom to top, exhibit different symmetries, i.e., $E_{2}$, $A_{3}$, $E_{1}$%
, $A_{2}$ $\cdots $, the corresponding valence band levels show $E_{1}$, $%
A_{4}$, $E_{2}$, $A_{1}$ $\cdots $. For zigzag GQDs with odd $N_{z}$, the
conduction band levels display the opposite (same) symmetries $E_{1}$, $%
A_{4} $, $E_{2}$, $A_{1}$, $\cdots $ to the conduction (valence) band levels
of zigzag GQDs with even $N_{z}$. For armchair GQDs, the lowest conduction
band level always shows the symmetries $E_{1}$, $A_{4}$, $A_{2}$, $E_{2}$, $%
\cdots $ from bottom to top and this order is independent of the size ($%
N_{a} $) of the armchair GQD.

\begin{figure}[tbp]
\includegraphics [width=\columnwidth]{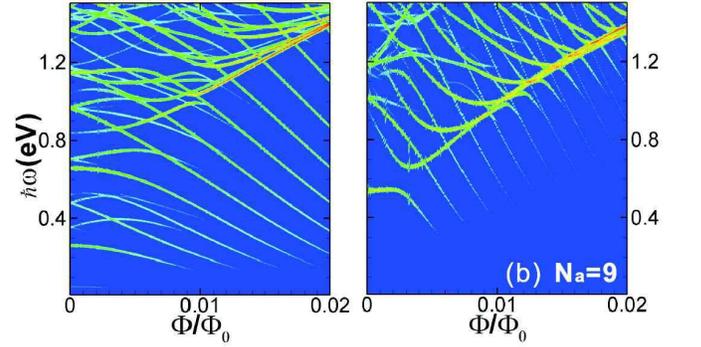}
\caption{ (Color online) The contour plot of the magneto-optical spectra of
zigzag (a) and armchair (b) GQD, respectively.}
\label{fig7}
\end{figure}

For $N_{z}=12$ ZGQD, the lowest optical-absorption peak (peak A) corresponds
to the transition between the lowest conduction band level $C_{1}$ with $%
E_{2}$ symmetry and the highest valence band level $V_{1}$ with $E_{1}$
symmetry. The second and third lowest transitions correspond to the
transition between the level $C_{2}\left( C_{3}\right) $ with $A_{3}\left(
E_{1}\right) $ symmetry and the level $V_{3}\left( V_{2}\right) $ with $%
E_{2}\left( A_{4}\right) $ symmetry and the level $C_{1}\left( C_{4}\right) $
and $V_{4}\left( V_{1}\right) $, respectively. But the strengths of these
three transitions are very small, therefore these transitions are not
clearly seen in the contour spectrum in Fig. \ref{fig7} at $\Phi /\Phi
_{0}=0 $. The strong absorption peak (peak D) appears at $E=0.26$ eV,
corresponding to the transition between the level $C_{3}$ with $E_{1}$
symmetry and the level $V_{3}$ with $E_{2}$ symmetry. This strong absorption
arises from the large moment matrix $\left\langle n\right\vert \mathbf{p}%
\left\vert m\right\rangle $ between these states.

For $N_{a}=9$ AGQD, the lowest peak (peak A) is similar to that in the
zigzag GQD, corresponding to the transition between $C_{1}$ and $V_{1}$. But
the second peak (peak B) is different from those of the zigzag GQD. This
peak corresponds to the transition between the level $C_{2}\left(
C_{4}\right) $ with $A_{4}\left( E_{2}\right) $ symmetry and the level $%
V_{4}\left( V_{2}\right) $ with $E_{1}\left( A_{3}\right) $ symmetry. The
third strong peak (peak C) indicates the transition between the level $C_{4}$
with $E_{2}$ symmetry and the level $V_{4}$ with $E_{1}$ symmetry. Strong
absorption takes place when the initial ($V_{i}$) and final states ($C_{i}$)
have either $E_{1}$ symmetry or $E_{2}$ symmetry. As the size of the GQD
increases, the absorption peaks shift to long wavelength for both ZGQD and
AGQD. The absorption peaks of the ZGQD shift to the long wavelength faster
than those of the AGQD. The relative strength between the peak D and A
increases as the size of the GQD increases for ZGQDs. But for AGQDs, the
relative strength between the peak C and A is almost independent of the size.

Next, we discuss the effect of a magnetic field on the optical spectrum of a
GQD. Here, we only focus on the small magnetic flux case (see Fig. \ref{fig7}%
). The spectra of two distinct GQDs, zigzag and armchair GQD,
exhibit quite different behavior due to their different level
structures and the oscillator strengths determined by the boundary,
especially for the LCL\ and HVL which localize at the edge of ZGQD.
The spectra of two distinct GQDs show that the strengths of the
transitions vary as the magnetic field increases. In particular, the
strong absorption lines exhibit $\sqrt{B}$ asymptotic\ behavior
corresponding to the transitions between the conduction and valence
band Landau levels at high magnetic field. We also find
anti-crossings in the spectra, since the magnetic field induces the
mixing of the levels belonging to the different families.

In summary, we investigated theoretically the magnetic levels and
the optical spectrum in GQDs. In contrast to conventional
semiconductor QDs, the LCL and HVL exhibit an edge-state feature,
i.e., a non-zero probability of being at the edge of the sample, and
the density distribution depends sensitively on the type of boundary
of GQDs and the magnetic field strength. The magnetic levels of GQD
display a Hoftstadter butterfly characteristic, and approach the
Landau levels of two-dimensional graphene as the magnetic field
increases. The magneto-optical spectrum of a graphene quantum dot in
the interesting energy range (0-3 eV) is promising for carbon-based
electronics applications. The position and strength of the
absorption peaks can be tuned by the size of the GQD, the type of
the edge of the GQD, and the external magnetic field.

\begin{acknowledgments}
This work is supported by the NSF of China Grant No. 60525405 and the
Flander-China bilateral programme.
\end{acknowledgments}

\end{document}